\documentclass[12pt]{article}
\usepackage{cite}
\usepackage{verbatim}
\usepackage{graphicx}
\usepackage{amssymb,graphicx}
\usepackage{epstopdf}
\usepackage{amsmath,amsfonts}
\usepackage{epsfig}

\def\d{\partial}
\def\l{\left(}
\def\r{\right)}

\newcommand{\avg}[1]{\left< #1 \right>} 
\newcommand{\be}{\begin{equation}}
\newcommand{\ee}{\end{equation}}
\newcommand{\bea}{\begin{eqnarray}}
\newcommand{\eea}{\end{eqnarray}}
\newcommand{\bg}{\begin{gather}}
\newcommand{\eg}{\end{gather}}
\newcommand{\bseq}{\begin{subequations}}
\newcommand{\eseq}{\end{subequations}}
\renewcommand{\ln}{\mathop{\rm ln}\nolimits}

\begin{document}

\begin{flushright}
\end{flushright}
\vspace{10pt}
\begin{center}
  {\LARGE \bf Cosmological phase transition, baryon asymmetry and dark matter Q-balls.} \\
\vspace{20pt}
E.~Krylov$^{a}$,~A.~Levin$^{a,b}$, V. Rubakov$^{a,b}$
\\
\vspace{15pt}
  $^a$\textit{
Lomonosov Moscow State University, Physics Faculty,\\
Department of Particle Physics and Cosmology.\\
GSP-1, Leninskie Gory, Moscow, 119991, Russian Federation
  }\\
  $^b$\textit{
Institute for Nuclear Research of
         the Russian Academy of Sciences,\\  60th October Anniversary
  Prospect, 7a, 117312 Moscow, Russia}\\
\vspace{5pt}
    \end{center}
    \vspace{5pt}

\begin{abstract}

We consider a mechanism of dark matter production in the course of first order phase transition.
We assume that there is an asymmetry between X- and $\bar{\text{X}}$-particles of dark sector.
In particular, it may be related to the baryon asymmetry.
We also assume that the phase transition is so strongly first order, that X-particles do not permeate into the new phase.
In this case, as the bubbles of old phase collapse, X-particles are packed into Q-balls with huge mass defect. These Q-balls compose the present dark matter.
We find that the required present dark matter density is obtained for the energy scale of the theory in the ballpark of 1--10 TeV.
As an example we consider a theory with effective potential of one-loop motivated form.

\end{abstract}
\section{Introduction.}
The idea that the baryon asymmetry and dark matter in the Universe may have common origin, 
is of considerable interest \cite{B1,B2,B3,B4,B5,B6,B7,B8,B9,B10,B11,B12,B13,B14,B15,B16,B17,B18,B19,B20}.
In particular, 
dark matter particles (X, $\bar {\text{X}}$) may have their own conserved charge and 
be created, together with baryons, in asymmetric decays of heavier 
paricles \cite{B3,B4,B5,B6,B7,B8,B15,B17,B20}. In this scenario, the initial asymmetry 
in the dark matter particles is roughly of the order of the baryon asymmetry,
\be 
\label{nXnq}
n_X-n_{\bar{X}} \sim n_B .
\ee
Assuming that X-$\bar {\text{X}}$ annihilation cross section is not particularly small, 
one obtains the estimate for the present dark matter mass density 
\be
\label{*}
\rho_X \sim m_X n_B.
\ee
Thus, the correct value of $\rho_X$ appears to require the mass of X-particles in a few GeV range.

New physics, whose manifestation would be 
the existence of X-particles,  may well be characterized by much higher energy scale, and 
the X-particle mass may grossly exceed a few GeV. One may wonder whether the 
co-production of the baryon asymmetry and dark matter can still work in this case. The 
estimate (\ref{*}) shows that heavy X-particles are overproduced, so one needs a 
mechanism that makes the actual mass density of dark matter much lower than 
that given by (\ref{*}).

In this paper we consider a possible scenario of this sort. 
The idea is that X-particles may be packed into Q-balls. 
A Q-ball made of X-particles of total number $Q$ typically has a mass $m_Q$, which is much smaller than 
$Q \cdot m_X$ \cite{FRS1,FRS2,FRS25,FRS3,coleman}. Hence, the mass 
density of the dark matter Q-balls is naturally well below the estimate (\ref{*}). 
A mechanism that packs free particles into Q-balls applies to the Friedberg--Lee--Sirlin 
Q-balls \cite{FRS1,FRS2,FRS25,FRS3} (as opposed to the Coleman Q-balls \cite{coleman} 
explored in supersymmetric theories \cite{susy1,susy2}) and is as follows \cite{HotBB}.

\begin{figure}[htb]
\label{F1}
\begin{center}
\includegraphics[width=\textwidth,height=1in]{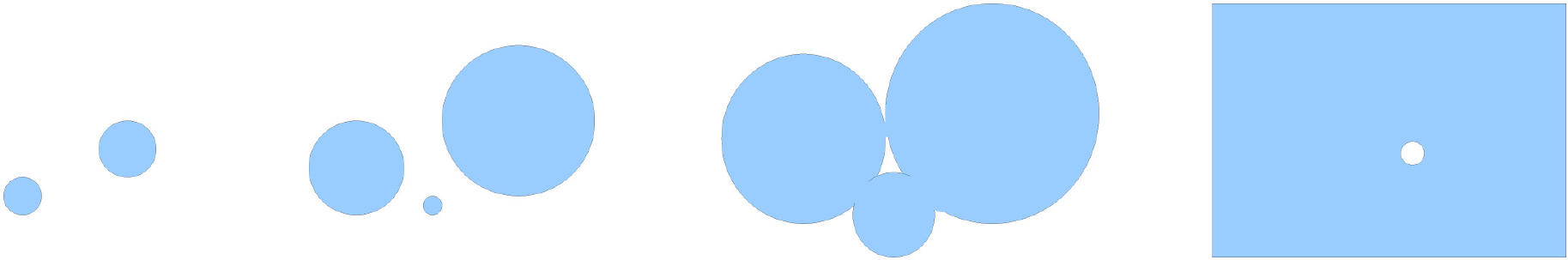}
\caption{\label{transition}Q-ball formation during the first-order phase transition 
from the phase $\phi=0$ (white) to the phase $\phi \ne 0$ (blue).}
\end{center}
\end{figure}

Let us assume that X-particles obtain their mass due to the interaction with an 
additional scalar field $\phi$, so that $m_X=h\phi$, where $h$ is the coupling 
constant. Let us also assume that there is the first order cosmological
phase transition at some temperature $T_c$ from the phase $\phi=0$ to the phase 
$\phi=\phi_c \ne 0$, and, furthemore, that the X-particle mass is large in the new phase, 
$h\phi_c \gg T_c$. Then X-particles get trapped in the remnants of the old phase, 
which eventually shrink to very small size and become Q-balls (see Fig. \ref{F1}).\\

We find that this mechanism indeed works in a certain range of couplings characterizing the model. Interestingly, 
the correct value of the present Q-ball mass density is obtained 
for X-particle mass in the ballpark of 1--10 TeV. This makes the scenario 
potentially testable in collider experiments. 
The Q-ball mass and charge are in the range 
$m_Q \sim 10^{-6}-10^{-3}$ g and $Q \sim 10^{19} - 10^{21}$, 
respectively. In this respect our Q-ball dark matter is not very different from 
that discussed in the context of supersymmetric theories \cite{susy1,susy2}.
Phenomenology of our dark matter Q-balls is also similar to that of supersymmetric 
Q-balls \cite{susy1,susy2}, except that the latter may be stabilized by baryon number 
and hence eat up baryons.

The paper is organized as follows. 
We begin with a brief description of Q-balls 
(Section \ref{Q-balls}).
In Section \ref{phasetr} we discuss the creation of Q-balls in the course of the phase transition and relate the dark matter Q-ball parameters and their present mass density to the properties of the phase transition.
We also consider the conditions of validity of our scenario.
In Section \ref{1loop} we give a concrete example based on one-loop motivated form of  the finite temperature effective potential and present the ranges of parameters for which our mechanism is viable. We conclude in Section \ref{conc}.

\section{Q-balls.}
\label{Q-balls}

\subsection{Q-ball configuration}
Q-balls are compact objects that exist in some models possessing a global symmetry and 
associated conserved charge. One of the simplest models admitting Q-balls is 
of the Friedberg--Lee--Sirlin type~\cite{FRS1,FRS2,FRS25,FRS3}. 
Its Lagrangian is\footnote{As it stands, this model has the discrete symmetry 
$\phi \to-\phi$ and hence is not viable because of the domain wall problem. To get around 
this problem one can either assume that the discrete symmetry is explicitly broken 
or consider the 
field $\phi$ carrying some gauge or global charge. This qualification is 
irrelevant for what follows.}
\begin{subequations}
\begin{align}
L = {1 \over 2}(\d_\mu \phi)^2 & - U(\phi)
+ ( \d_\mu \chi)^{*}(\d_\mu \chi) - h^2 \phi^2 \chi^{*} \chi \; ,
\label{L}
\\
U(\phi) &=\lambda (\phi^2-v^2 )^2 \; ,
\label{may14-12-1}
\end{align}
\end{subequations}
where $\phi$ is a real scalar field and $\chi$ is a
complex scalar field. The field $\chi$ is meant to describe 
X-particles whose mass in vacuo equals 
\[
m_X=hv \; .
\]
These particles carry global charge, associated with
the $U(1)$-symmetry $\chi \to \mbox{e}^{i\alpha}
\chi$. The lowest energy state of large enough charge is a spherical Q-ball with
$\phi=0$ inside and $\phi=v$ outside. At large $Q$,
its size $R$ and energy $E$
are determined by the balance of the energy of $Q$ massless
$\chi$-quanta confined in the potential well of radius $R$ and 
the potential energy of the field $\phi$ in the interior, i.e., 
$R$ and $E$  are
found by
minimizing
\be
E(R) = \frac{\pi Q}{R} + \frac{4\pi}{3} R^3 U_0 \; ,
\label{2.1}
\ee
where $U_0 = U(0) - U(v) = \lambda v^4 $. Hence, the Q-ball parameters are
\be
R_Q = \l \frac{Q}{4U_0}\r^{1/4}, \ \ \ 
m_Q = \frac{4\sqrt{2} \pi}{3} Q^{3/4} U_0^{1/4} \; .
\label{ER}
\ee
Note that the surface energy is proportional to $R_Q^2 \propto Q^{1/2}$
and therefore negligible at large $Q$. 
The Q-ball is stable provided its energy is smaller than the rest energy
of $Q$ massive $\chi$-quanta in the vacuum $\phi=v$,
\be
\label{stabil}
m_Q < m_X Q \; .
\ee
A Q-ball is a 
classical object, since its raduis is much larger than its Compton wavelength.

\subsection{Radius of cosmological Q-balls}
Assuming that Q-balls made of X-particles compose dark matter and that the 
X-asymmetry is related to the baryon asymmetry via Eq.~(\ref{*}), 
we can obtain an estimate for the present radius of a typical Q-ball already 
at this point. Indeed, we find from Eq. (\ref{ER}) that
\[
R_Q= 4\pi \frac{Q}{3m_Q} \; .
\]
Now, since X-particles are packed into Q-balls, we have
\[
\frac{(n_X-n_{\bar{\text{X}}})}{s}=\frac{n_Q Q}{s} \; ,
\]
where $n_Q$ is the number density of Q-balls and $s$ is the entropy density.
Therefore
\be
\frac{n_Q Q}{s} = \frac{n_q-n_{\bar q}}{s} = 3 \Delta_B \; ,
\label{may16-1}
\ee
where $\Delta_B=n_B/s=0.9\times10^{-10}$, and we assume for 
definiteness that the  X-particle asymmetry is equal to the
quark asymmetry. Making use of the relation $\rho_{DM}=m_Q n_Q$, we obtain
\be
\left. \frac{Q}{m_Q} \right|_{t_0}= \frac{3 \Delta_B s_0}{\rho_{DM}},
\ee
where the subscript "0" refers to the present epoch. Hence, we obtain from Eq. \eqref{ER} that the present radius is
\be
\label{RQ}
R_Q = 4\pi \frac{\Delta _Bs_0}{\rho _{DM}} \simeq 6 
\cdot 10^{-14}~\text{cm} .
\ee
Thus, even though the Q-ball mass and charge depend on the
parameters of the model, its typical radius does not.

\section{Q-ball production at the cosmological phase transition.}
\label{phasetr}

In this section we give a general description of the Q-ball formation without specifying the form of the finite temperature effective potential. We assume only that it has a minimum $\phi =0$ at high temperature, and that the phase transition from $\phi = 0$ to $\phi \neq 0$ is of the first order.

\subsection{Bubble nucleation rate}

Let $T_c$ be the critical temperature at which the effective potential has two degenerate minima at $\phi=0$ and $\phi=\phi_c$. Below this temperature, the new phase $\phi=\phi_c$ has lower free energy density,
\be 
V_{eff}(\phi_c)-V_{eff}(0)=-\rho<0.
\ee
Thermal fluctuations lead to the creation of the bubbles of the new phase. 
In the thin wall approximation, the free energy of a bubble of radius $R$ is
\be 
\label{F(R)}
F(R)=-{4 \over 3}\pi R^3\rho+4\pi R^2\sigma \; ,
\ee
where $\sigma$ is surface free energy density.
The extremum of (\ref{F(R)}) gives the free energy of the critical bubble,
\be 
F_c=\frac{16\pi}{3} \frac{\sigma^3}{\rho^2} \; . 
\ee
It, in turn, determines the bubble nucleation rate,
\[
\Gamma=\kappa T_c^4 e^{-\frac{F_c(T)}{T_c}},
\]
where $\kappa$ is a factor roughly of order 1.
Let us introduce 
\be
\tau= \frac{T_c-T}{T_c} \; ;
\ee 
and assume that this parameter is small. 
To the leading order in $\tau$, we have
\begin{align} 
\label{rho1}
\rho &=\rho_1 \tau \; ,
\\ 
\label{sigma0}
\sigma &= \sigma_0,
\end{align}
where $\rho_1$ and $\sigma_0$ are constants independent of $\tau$.
Thus
\be
\Gamma=\kappa T_c^4 e^{-\frac{A}{\tau^2}},
\label{may15-1}
\ee
where
\be 
A=\frac{16\pi}{3} \frac{\sigma_0^3}{\rho_1^2}. 
\label{may16-5}
\ee
As the temperature decreases, the bubble nucleation rate rapidly grows.

\subsection{Temperature of the phase transition}
\label{Dyn}

To estimate the number density and charge of Q-balls produced during the
phase transition, we need an estimate of the transition
temperature. Since the duration of the phase transition is often
much shorter than the Hubble time (this corresponds to $\tau \ll 1$),
we neglect the cosmological expansion and write 
for the fraction of volume occupied by the old phase at time $t$ \cite{Guth-Weinberg}
\[
x(t)=\exp[-\Delta(t)] \; ,
\]
where
\[
\Delta(t)=\int_{t_c}^t V(t,t') \Gamma(t')  dt' \; ,
\]
$V(t,t') = \frac{4\pi}{3} \left[ u(t-t^\prime)\right]^3$ 
is the volume, at time $t$,
 of a bubble of the new phase born at time $t'$, and $u$ is the
velocity of the bubble wall. 
We make use of the standard relation between the Hubble parameter and temperature,
$H= T^2/M_{Pl}^*$, where
$M_{Pl}^*=\frac{M_{Pl}}{1.66\sqrt{g_*}}$ and 
$g_*$ hereafter denotes the effective number of the degrees of freedom at the phase transition temperature.
We obtain for $\tau \ll 1$
\be 
\Delta =\kappa u^3 \left( \frac{M_{Pl}^*}{T_c}\right)^4 \int_{0}^\tau
\left( \tau-\tau'\right)^3 \mbox{e}^{-\frac{A}{\tau'^2}}
d\tau' \; .
\label{intDelta}
\ee
This integral is saturated near the upper limit of integration, and we get
\[
\Delta \sim \kappa u^3 \left( \frac{M_{Pl}^*}{T_c}\right)^4 
\frac{\tau^{12}}{A^4} \mbox{e}^{-\frac{A}{\tau^2}} \; . 
\]
The phase transition occurs when $\Delta$ is roughly of order 1, 
which happens
in a narrow interval of temperatures around 
\be
\tau_*=A^{1/2}L^{-1/2} \; ,
\label{may15-6}
\ee
where
\be
L  = \ln \left[ \kappa u^3 A^2 \left( \frac{M_{Pl}^*}{T_c}\right)^4 \right] \; ,
\label{may16-7}
\ee
with logarithmic accuracy.
Our estimate is valid provided that $\tau_* \ll 1$, i.e.
\be
\label{Acond}
A \ll L \; .
\ee
In what follows we assume that this is indeed the case, see also Section \ref{cond}.

Note that at the time of the phase transition, the bubble nucleation rate 
\eqref{may15-1}
is
still small,
\be
\Gamma \sim \frac{T_c^4 A^4}{u^3 \tau_*^{12}}\left( \frac{T_c}{M_{Pl}^*}\right)^4 .
\label{may15-3}
\ee
This is of course a consequence of the slow cosmological expansion.

\subsection{Q-balls in the end of the phase transition}

We are now ready to estimate the volume from which X-particles are collected
into a single Q-ball. This volume will determine the number denisty 
of Q-balls immediately after the phase transition
and the typical Q-ball charge. One way to obtain the estimate is to notice that
within a factor of order 1, this volume is the same as the volume of
a remnant of the old phase in the midst of the phase transition, see 
Fig.~\ref{transition}. We estimate
the size $R_*$ of a remnant by requiring that it shrinks to small size
before a bubble of the new phase is created inside it.
The lifetime of a
remnant of size $R$ is $R/u$, so the latter requirement gives
\[
R_*^3\Gamma(T)\frac{R_*}{u} \sim 1 \; .
\]
Making use of Eq. \eqref{may15-3}
we obtain
\be 
R_* \sim \frac{u\tau_*^3M_{Pl}^*}{T_c^2 A}.
\label{may15-5}
\ee
Another way to estimate the volume that will shrink to one Q-ball is to consider
bubbles of the new phase instead, and estimate the typical size of a bubble in the
midst of the transition. At time $t$, the average bubble volume is
\[
\frac{4\pi}{3} R^3 (t) = N^{-1}(t)
\int_{t_c}^t~V(t,t^\prime) \Gamma (t') x(t')~dt' \; ,
\]
where $N(t) = \int_{t_c}^t~\Gamma (t') x(t')~dt'$ estimates the number density of the bubbles.
These integrals are evaluated in the same way as Eq. \eqref{intDelta}, and we get
\[
\frac{4\pi}{3} R_*^3 (t) =2\pi  \l\frac{u\tau_*^3 M_{Pl}^*}{T_c^2 A}\r^3 \; .
\]
This gives the same estimate as \eqref{may15-5}, which demonstrates the consistency
of the approach.

From now on we use the following expression for the  
volume from which X-particles are collected
into a single Q-ball:
\[
V_* = \frac{4\pi}{3} R_*^3=\xi \l \frac{u A^{1/2} M_{Pl}^*}{T_c^2 L^{3/2}} \r^3 \; ,
\]
where we inserted $\tau_*$ given by Eq.~\eqref{may15-6}, and
$\xi$ is a parameter of order 1 which parmetrizes the uncertainty of our estimate.

The number density of Q-balls immediately after the phase transition equals
$n_Q (T_c) = V_*^{-1}$, and its ratio to the entropy density is
\be
\frac{n_Q (T_c)}{s (T_c)} = \frac{45}{2\pi^2 g_*} \frac{1}{T_c^3 V_*} \; .
\label{may16-3}
\ee
Note that for given values of couplings, this ratio is suppressed by
$(T_c/M_{Pl})^3$. Since one Q-ball contains all excess of X-particles in
volume $V_*$, its charge is (again assuming $X$-asymmetry equal to quark
asymmetry, cf. Eq.~\eqref{may16-1})
\be
Q = n_X V_* = 3 \Delta_B s (T_c) V_* \; .
\label{may16-2}
\ee
This charge is large, since it is proportional to $(M_{Pl}/T_c)^3$.
The X-particles are packed into Q-balls rather efficiently.

\subsection{Q-balls at present.}
\label{now}
Once the phase transition completes and Q-balls get formed, the ratio of
their
number density to entropy density stays constant
and given by Eq.~\eqref{may16-3}.
With the Q-ball charge \eqref{may16-2}, its mass is found from Eq.~\eqref{ER},
\[
m_Q =\frac {4\pi \sqrt2 }{3} {U_0}^{1/4} \left[3\Delta_B s(T_c)\right]^{3/4} {V_*}^{3/4}=
\]
\be
=
7.3\cdot\xi^{3/4} \cdot 
{\Delta_B}^{3/4} {g_*}^{3/4} {M^*_{Pl}}^{9/4} 
\frac{u^{9/4} U_0^{1/4} A^{9/8}}{T_c^{9/4} L^{27/8}},
\label{mQ}
\ee
Hence, Q-balls are dark matter candidates provided that
\be
\frac{m_Q n_Q}{s} = f \Delta _B^{3/4}\frac{ T_c^{3/4} U_0^{1/4} }{ {M^*_{Pl}}^{3/4} }= \frac{\rho_{DM}}{s_0} ,
\label{may16-10}
\ee
where $\rho_{DM} \simeq 1 \cdot 10^{-6}~\mbox{GeV} \cdot \mbox{cm}^{-3}$ and 
$s_0 \simeq 3000~\mbox{cm}^{-3}$ are the present mass density of dark matter
and entropy density, respectively, and
\be
\label{f}
f = 19.0    
\frac{\xi ^{-1/4} L^{9/8} }{A^{3/8} g_*^{1/4}\text{  }u^{3/4} } ,
\ee
is a combination of dimensionless parameters.
As we see, the dependence on parameter $\xi$ is weak, so we set $\xi=1$ from now on.
At this point we can make rough estimate for the relevant energy scale.
Assuming $U_0\sim T_c^4$, $A\sim 1$, $u \sim 0.1$, $g_* \sim 100$ and $L \sim 100$, we get from Eq. \eqref{may16-10}  that $T_c$ should be in a ballpark of 
\be
\label{scale}
T_c \sim 1 \div 10~\text{TeV}.
\ee
As we pointed out in  Introduction, 
this relatively low energy scale makes the scenario interesting from the viewpoint of collider experiments.
 
\subsection{Validity of calculation.}
\label{validity}
A particle physics model in which our mechanism can work must satisfy several requirements. 
One is the condition \eqref{Acond} or
\be
\label{cond1}
A \ll 4\text{ln} \l \frac{M^*_{Pl}}{T_c} \r .
\ee
Another is that X-particles do not penetrate into the new phase in the course of the phase transition.
This is the case if their mass in the new phase is sufficiently larger than temperature. 
Quantitatively, we require that the mass density of remaining free X-particles is negligible compared to the mass density of dark matter Q-balls,
\be
\label{rhonas}
\frac{ (n_X + n_{\bar{X}}) \left. m_X \right|_{(T=0)}}{s} \ll \frac{\rho_{DM}}{s_0}.
\ee
Let us see what this condition means in terms of parameters.

The dynamics of penetration of X-particles into the new phase depends on the bubble wall velocity and the strength of X-particle interations with cosmic plasma. There are two extreme cases: very slow motion of the bubble wall and very fast motion. Let us consider them in turn.

{\bf Slow wall.}

Let the wall velocity be so small that there is
complete thermal equilibrium for X-particles across the wall. Then 
the chemical potentials in the old and new phase are equal, and since
X-particles in the old phase are massless, the chemical potential is
negligibly small: indeed, in the old phase
$\mu T^2 \sim (n_X - n_{\bar{X}}) \sim \Delta_B T^3$, hence $\mu/T \sim \Delta_B$.
The number
density of X-particles in the new phase is given by the equilibrium formula
for non-relativistic species, and we have
\be
\frac{n_X + n_{\bar{X}}}{s} = s^{-1} \cdot 2 \l \frac{m_X T}{2\pi} \r^{3/2}
\mbox{e}^{-m_X/T} \sim \frac{1}{g_*}  \l \frac{m_X }{2\pi T} \r^{3/2}
\mbox{e}^{-m_X/T}\; ,
\label{may16-b1}
\ee
where all quantities, including $m_X$, are evaluated at $T_c$.

{\bf Fast wall.}

In the opposite case of fast wall, all X-particles that penetrate the new
phase stay there. Flux of X-particles and $\bar{\text{X}}$-particles,
whose momentum normal to the wall
exceeds $m_X$ is
\[
j_{X\bar{X}} =  \frac{2}{(2\pi)^3} \int~d^2 p_{L} \int_{m_X}^\infty d p_T 
\mbox{e}^{-\sqrt{p_T^2 + p_L^2}/T} \; .
\]
This integral is saturated at $p_T$ near $m_X$ and $p_L \ll p_T$. Hence, we write
\[
\sqrt{p_T^2 + p_L^2} =
 p_T +  \frac{p_L^2}{2p_T}
= p_T + \frac{p_L^2}{2m_X}
\]
and  obtain
\[
j_{X\bar{X}} =  \frac{1}{2\pi^2} m_X T^2 \mbox{e}^{-m_X/T} .
\]
As the wall moves, its radius increases by $dR = u dt$, and
the number of pernetrated particles is 
\[
j_{X\bar{X}}S dt  = j_{X\bar{X}} u^{-1} 4\pi R^2 dR.
\]
So, the number density in the new phase is in the end
\[
n_X + n_{\bar{X}} = u^{-1} j_{X\bar{X}} .
\]
This gives the estimate similar to that in the slow wall case, except that
instead of particle velocity
$\sqrt{T/ m_X}$ it involves wall velocity.

If X$\bar{\text{X}}$-annihilation is switched off in the new phase,
then the ratio $(n_X + n_{\bar{X}})/s$ stays constant after the phase transition.
For any of the above cases, and hence for all intermideate ones, up to logarithmic corrections we get from Eq. \eqref{rhonas}
\be
\frac{\left. m_X \right|_{T_c}}{T_c} >
\text{ln} \l \frac{\rho_{DM}}{s_0 T_c} \r ,
\ee
or
\be
\label{hphi}
\frac{h \phi_c }{T_c} >
\text{ln} \l \frac{3\cdot 10^{-10} \text{GeV}}{T_c} \r.
\ee
This condition does not apply if X$\bar{\text{X}}$-annihilation is efficient in the new phase.
If Eq. \eqref{hphi} does not hold, the behavior of X-particles in the new phase is similar to that of WIMPs (note that the X-particle abundance in the new phase just after the phase transition either coincides with, or exceeds the equilibrium abundance). In that case the condition \eqref{rhonas} implies that the X$\bar{\text{X}}$-annihilation cross section exceeds the standard WIMP annihilation cross section. 
Since the energy scale inherent in the model is high (see Eq. \eqref{scale}) the latter scenario is not particularly plausible.
In what follows we assume that the inequality \eqref{hphi} holds.
 
\section{Example. One-loop motivated effective potential.}
\label{1loop}
\subsection{Critical temperature and Q-ball parameters.}
As an example, let us consider effective potential of a particular, one-loop motivated form,
\be 
\label{Veff}
U(\phi,T)=\alpha(T^2-T_{c2}^2)\phi^2-\gamma T \phi^3+\lambda\phi^4,
\ee
where $T_{c2}$, $\alpha$ and $\gamma$ 
are parameters depending on particle physics at temperature $T$, and we
neglect temperature corrections to the quartic self-coupling.
We treat  $T_{c2}$, $\alpha$, $\gamma$ and $\lambda$ as parameters of the model, but
keep in mind the relation $\alpha T_{c2}^2 = 2\lambda v^2$, which
follows from Eq. \eqref{may14-12-1}.

We assume in what follows that $9\gamma^2 < 32 \alpha \lambda$. 
Then at high temperatures, the effective potential has only one local minimum at $\phi=0$. 
As the Universe cools down, the minimum at $\phi \ne 0$ develops. It becomes deeper than the minimum at $\phi=0$ at $T<T_c$, where
\be
\label{TcTc2}
T_c^2= \frac{4\alpha \lambda}{4\alpha \lambda - \gamma^2} T_{c2}^2 \; . 
\ee
In what follows we need the expression for 
the height of the scalar potential and mass of the $\phi$-particle, both
at zero temperature, 
in terms of the parameters entering (\ref{Veff}) and (\ref{TcTc2}), 
\be
U_0 = \frac{(4\alpha \lambda -\gamma^2)^2 }{64\lambda^3}T_c^4 \; , \;\;
\;\;\;\;\; 
m_\phi = \l \frac{4\alpha \lambda - \gamma^2}{\lambda}\r^{1/2} T_c \; .
\label{may14-3}
\ee
In this way, we trade the parameter $v$ in the original Lagrangian (\ref{L}) 
for the parameters relevant for the phase transition.

Let us now rewrite the results of Section \ref{phasetr}
 in terms of the parameters used in (\ref{Veff}).
At small $\tau \equiv \frac{T_c-T}{T_c}$ one has
\begin{align} 
\label{rho}
\rho &=-\frac{ \gamma ^2 \left(4 \alpha  \lambda -\gamma ^2\right)  }{8 \lambda ^3} T_c^4 \tau \; ,
\\ 
\label{sigma}
\sigma &=
\int_0^{\phi_c}\sqrt{2U(\phi,T_c)} d\phi= \frac{\gamma^3}{24 \sqrt{2} \lambda^{5\over 2}}T_c^3.
\end{align}
Thus
\be
\label{Aform}
A=\frac{\pi  }{81 \sqrt{2} }\frac{\gamma ^5}{ \lambda ^{3/2} \left(4 \alpha  \lambda  - \gamma^2\right)^2}
\ee
and
\be
V_* =  
 \l \frac{u A^{1/2} M_{Pl}^*}{T_c^2 L^{3/2}} \r^3=
 \l  \frac{\sqrt{\pi} u} {9\cdot2^{1/4} }  \frac{ M^*_{Pl}}{T_c^2} \frac{\gamma^{5/2}} {\lambda^{3/4}(4\alpha \lambda - \gamma^2)}\times L^{-3/2} \r^3,
\ee
where $L$ 
is defined in Eq. 
\eqref{may16-7}.
We find from Eq. \eqref{may16-10} the present Q-ball dark matter mass density
\be
\label{rhoQ}
\rho_{DM} = K_{\rho } \Delta _B^{3/4}s_0 {M^*_{Pl}}^{-3/4} g_*^{-1/4}u^{-3/4} \gamma ^{-15/8} \lambda ^{-3/16} \left(4 \alpha  \lambda -\gamma ^2\right)^{5/4}T_c^{7/4}L^{9/8},
\ee
where
\[
K_{\rho }=2^{15/16}3^{7/4} 5^{1/4} \pi ^{1/8}\approx 22.5.
\]
Equating it to the actual dark matter density we obtain the critical temperature in terms of other parameters,
\be
\label{Tcfin}
T_c= K_T\rho _{DM}^{4/7}\Delta _B^{-3/7}s_0^{-4/7}{M^*_{Pl}}^{3/7}g_*^{1/7}u^{3/7}\gamma ^{15/14}\lambda ^{3/28}\left(4\alpha \lambda -\gamma ^2\right)^{-5/7}\times L^{-9/14},
\ee
where 
\[
K_T=2^{-15/28}3^{-1}5^{-1/7}\pi ^{-1/14}\approx 0.17.
\]
In further computations we use $T_c=10$ TeV in the argument of logarithm \eqref{may16-7}, see Eq. \eqref{scale}.
Finally, Eqs. \eqref{may16-2} and \eqref{mQ} give for the Q-ball parameters
\be
Q= K_Q \rho _{DM}^{-12/7} \Delta _B^{16/7}s_0^{12/7}{M^*_{Pl}}^{12/7}{g_*}^{4/7}\text{  }\gamma ^{30/7}\text{    }u^{12/7}\text{  }\alpha ^{29/7} \lambda ^{9/7}\left(4\alpha \lambda -\gamma ^2\right)^{-5}L^{-18/7},
\ee
\be
m_Q=K_{m_Q}\rho _{DM}^{-5/7} \Delta _B^{9/7}s_0^{5/7} {M^*_{Pl}}^{12/7}{g_*}^{4/7}u^{12/7}\gamma ^{30/7}\lambda ^{-18/7}\left(4 \alpha \lambda -\gamma ^2\right)^{-6/7}L^{-18/7},
\ee
where
\[
K_Q=2^{69/7}\pi ^{26/7}3^{-4} 5^{-4/7}\approx 0.063,
\]
\[
K_{m_Q}=2^{13/7} 3^{-5} 5^{-4/7}\pi ^{26/7}\approx 0.42.
\]
The present number density of Q-balls is of course equal to $\rho_{DM}/m_Q$.

\subsection{Parameter space.}
\label{cond}
\begin{figure}[htb]
\begin{center}
\includegraphics[width=\textwidth,height=3.5in]{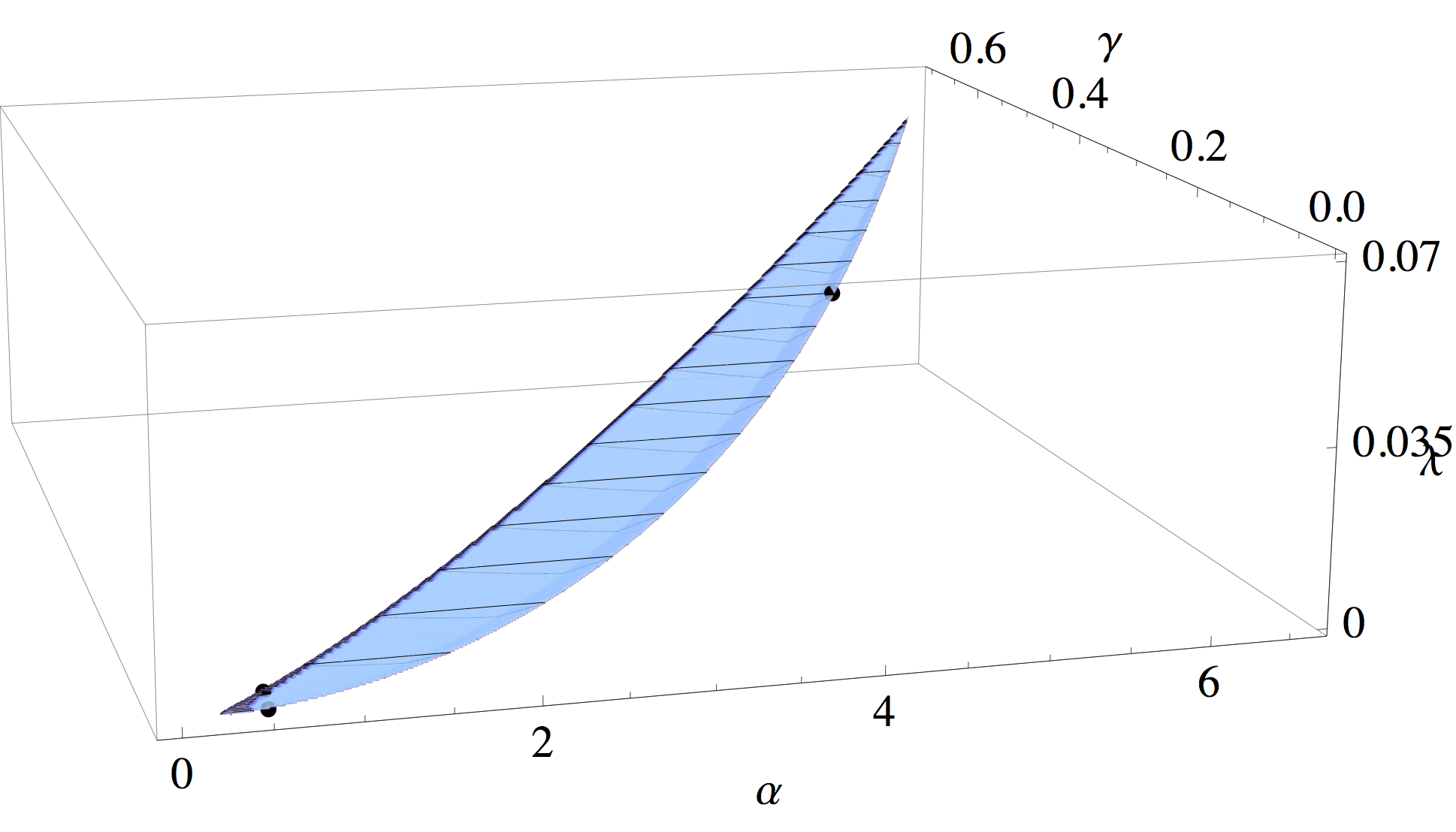}
\caption{\label{F2}
The region in the parameter space consistent with the constraints \eqref{Acond1}, \eqref{25}, \eqref{64pi} for $h=5$. Dots are the points for which the mass, critical temperature and Q-ball parameters are listed in Tables 1 and 2.}
\end{center}
\end{figure}

\begin{figure}[htb]
\begin{center}
\includegraphics[width=\textwidth,height=3.5in]{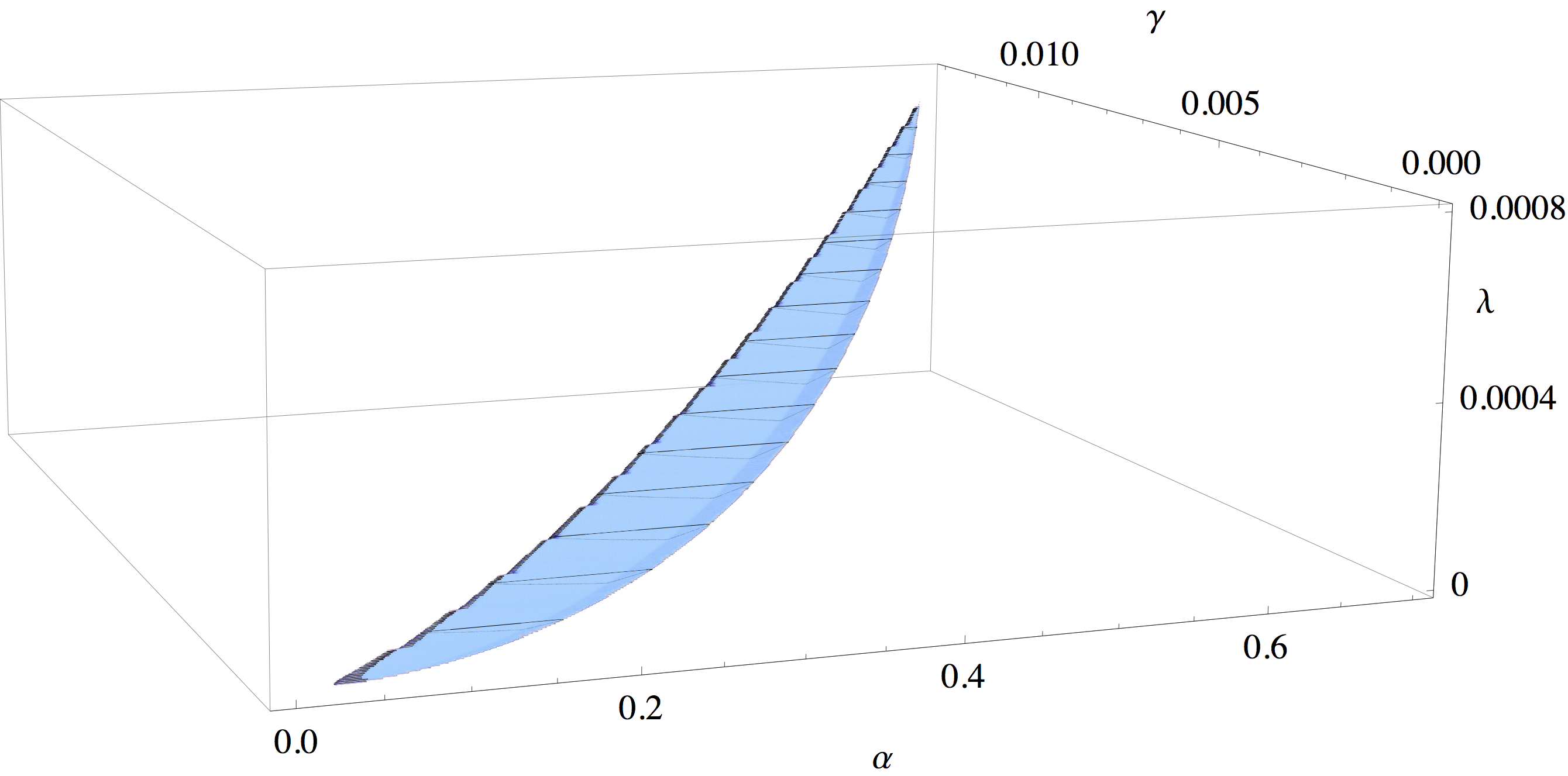}
\caption{\label{F3}
Same as in Fig. \ref{F2}, but for $h=3$.}
\end{center}
\end{figure}

In Section \ref{validity} we pointed out two conditions that the model should obey.
In terms of the parameters of the effective potential the condition \eqref{cond1} takes the following form (see Eq.\eqref{Aform})
\be
\label{Acond1}
\frac{\pi  }{81 \sqrt{2} }\frac{\gamma ^5}{ \lambda ^{3/2} \left(4 \alpha  \lambda  - \gamma^2\right)^2} \ll
4 \text{ln} \l \frac{M^*_{Pl}}{T_c} \r \simeq 120,
\ee
The concrete form of the condition \eqref{hphi} is obtained by noticing that
\be
\phi_c\equiv\left.\avg{\phi} \right|_{T=T_c}=\frac{\gamma}{2\lambda} T_c~.
\ee
We find
\be
\label{25}
h \frac{\gamma}{2 \lambda} > 25 .
\ee
We also assume that the quartic self-coupling of the field $\phi$ is not particularly small, and impose a mild constraint motivated by naturalness argument,
\be
\label{64pi}
\lambda>\frac{\alpha^2}{64\pi^2}
\ee
The conditions \eqref{Acond1}, \eqref{25} and \eqref{64pi} are actually quite restrictive.
In particular, they require that the $\phi$-$\chi$ coupling is rather strong. For $h \sim 5$, the available region in the parameter space is fairly large, as shown in Fig. \ref{F2}. This region becomes considerably smaller already for $h \sim 3$, see Fig. \ref{F3}.

\newpage
 We scanned the available regions and found the following range of masses $m_\phi$ and critical temperatures, at which our mechanism does work
\[
m_\phi \in  3\cdot 10^2 
\div 1 \cdot 10^4 \text{ GeV}, \]\[
T_c \in   1\cdot 10^3
 \div 3\cdot10^4 \text{ GeV} .
\]
For the Q-ball parameters we obtained the range
\[
m_Q \in   3\cdot10^{-7}  \div  3\cdot10^{-3} \text{ gram} , \]\[
Q \in  10^{19} \div  10^{22} ,\]\[
n_Q \in 1\cdot 10^{-27} \div 3\cdot 10^{-24} \text{ cm}^{-3}.
\] 
Three particular examples, corresponding to the points in Fig. \ref{F2}, are listed in Tables 1 and 2 for two values of the bubble wall velocity.
\newpage

{\bf Table 1.} Q-ball parameters for particular values of couplings (three left columns) and bubble wall velocity $u=0.03$.\label{table1}
\[
\begin{array}{|c|c|c|c|c|c|c|c|} \hline
 &    &  & m_\phi~\text{,GeV} & T_c\text{,GeV} & \text{Q} & m_Q\text{,gram} & n_Q,\text{cm}^{-3} \\ \hline
\alpha=5   & \gamma=0.4 & \lambda=0.04 & 1.9\times 10^4 & 4.8\times 10^3 & 8.7\times 10^{19} & 4.2\times 10^{-5} & 4.5\times 10^{-26}  \\ \hline
\alpha=0.5   & \gamma=0.02 & \lambda=0.002 & 8.1\times 10^3 & 6.1\times 10^3 & 7.2\times 10^{19} &2.3 \times 10^{-5} & 8.1\times 10^{-26}  \\ \hline
\alpha=0.5 & \gamma=0.004 & \lambda=0.0004 &  3.7\times 10^3 & 2.6\times 10^3 & 1.9\times 10^{19} & 5.6\times 10^{-6} & 3.4\times 10^{-25} \\ \hline
\end{array}
\]

{\bf Table 2.} Same but for $u=0.3$.\label{table2}
\[
\begin{array}{|c|c|c|c|c|c|c|c|} \hline
 &   &  & m_\phi~\text{,GeV} & T_c\text{,GeV} & \text{Q} & m_Q\text{,gram} & n_Q,\text{cm}^{-3} \\ \hline
\alpha=5   & \gamma=0.4 & \lambda=0.04 & 4.9\times 10^4 & 1.2\times 10^4 & 3.9\times 10^{21} & 1.9\times 10^{-3} & 1.0\times 10^{-27} \\ \hline
\alpha=0.5   & \gamma=0.02& \lambda=0.002 & 2.1\times 10^4 & 1.6\times 10^4 & 3.2\times 10^{21} & 1.0\times 10^{-3} & 1.9\times 10^{-27} \\ \hline
\alpha=0.5 & \gamma=0.004 & \lambda=0.0004  & 9.4\times 10^3 & 6.7\times 10^3 & 8.2\times 10^{20} & 2.5\times 10^{-4} & 7.8\times 10^{-27} \\ \hline
\end{array}
\]

\section{Conclusion.}
\label{conc}
We considered a mechanism producing Q-balls in the course of the first order phase transition and described it quantitatively.
As we have seen, this mechanism efficiently packs massive stable particles, which otherwise would be overproduced, and thus drastically reduces the mass density.

Using a well-studied model of Friedberg--Lee--Sirlin Q-balls, we obtained formulae for the dark matter properties, such as charge, mass and concentration of dark matter Q-balls, as well as masses of scalar fields, depending on the parameters of the transition.
As an example, we considered this mechanism in a theory with the effective potential of one-loop motivated form.

We have seen that the main parameter is the temperature of the phase transition, whose adjustment yields the right value of the present dark matter density.
A remarkable property of the mechanism is that for a wide range of parameters, independently of particle physics and effective potential model, the estimate for the energy scale is $T_c \sim 1$--10 TeV.

The main requirement for effective packing of particles into Q-balls is that the phase transition is strongly first order. For one-loop motivated effective potential this implies strong $\phi$-$\chi$ coupling and gives the main constraint on available parameter space. However, for other models this is not necessaraly the case, since there are other ways to make the first order phase transition strong enough (see, e.g., \cite{phtr,demgor} and references therein).

Our study was motivated by the possibility that X-particle asymmetry is related to the baryon asymmetry. However, this is optional. The mechanism described works in the same way, with $3 \Delta_B$ replaced by $\Delta_X$, if $n_X$ is considered as yet another free parameter.

\section{Grants}

This work has been supported in part by the grant of the President of the Russian Federation NS-5590.2012.2, by the grant of the Ministry of Science and Education 8412 and by the grant RFBR 12-02-00653.


\begin{thebibliography}{99}
\bibitem{B1}
H. Davoudiasl, R. N. Mohapatra,
arXiv:1203.1247v1 [hep-ph].


\bibitem{B2}
J. McDonald,
arXiv:1201.3124v1 [hep-ph]
\bibitem{B3}
  L.~Covi, E.~Roulet and F.~Vissani,
  Phys.\ Lett.\ B {\bf 384} (1996) 169
  [hep-ph/9605319].
\bibitem{B4}
  L.~Covi and E.~Roulet,
  Phys.\ Lett.\ B {\bf 399} (1997) 113
  [hep-ph/9611425].
\bibitem{B5}
  M.~Plumacher,
  hep-ph/9807557.
\bibitem{B6}
  S.~D.~Thomas,
  Phys.\ Lett.\ B {\bf 356} (1995) 256
  [hep-ph/9506274].
\bibitem{B7}
  W.~Buchmuller, K.~Schmitz and G.~Vertongen,
  Nucl.\ Phys.\ B {\bf 851} (2011) 481
  [arXiv:1104.2750 [hep-ph]].
\bibitem{B8}
  N.~J.~Poplawski,
  Phys.\ Rev.\ D {\bf 83} (2011) 084033
  [arXiv:1101.4012 [gr-qc]].
\bibitem{B9}
  S.~M.~Barr,
  Phys.\ Rev.\ D {\bf 85} (2012) 013001
  [arXiv:1109.2562 [hep-ph]].
\bibitem{B10}
  H.~An, S.~-L.~Chen, R.~N.~Mohapatra and Y.~Zhang,
  JHEP {\bf 1003} (2010) 124
  [arXiv:0911.4463 [hep-ph]].
\bibitem{B11}
  N.~Haba and S.~Matsumoto,
  Prog.\ Theor.\ Phys.\  {\bf 125} (2011) 1311
  [arXiv:1008.2487 [hep-ph]].
\bibitem{B12}
  L.~J.~Hall, J.~March-Russell and S.~M.~West,
  arXiv:1010.0245 [hep-ph].
\bibitem{B13}
  K.~Petraki, M.~Trodden and R.~R.~Volkas,
  JCAP {\bf 1202} (2012) 044
  [arXiv:1111.4786 [hep-ph]].
\bibitem{B14}
  D.~S.~Gorbunov and A.~G.~Panin,
  Phys.\ Lett.\ B {\bf 700} (2011) 157
  [arXiv:1009.2448 [hep-ph]].
\bibitem{B15}
  R.~Kitano and I.~Low,
  Phys.\ Rev.\ D {\bf 71} (2005) 023510
  [hep-ph/0411133].
\bibitem{B16}
  G.~R.~Farrar and G.~Zaharijas,
  Phys.\ Rev.\ Lett.\  {\bf 96} (2006) 041302
  [hep-ph/0510079].
\bibitem{B17}
  D.~Suematsu,
  JCAP {\bf 0601} (2006) 026
  [astro-ph/0511742].
\bibitem{B18}
  V.~A.~Kuzmin,
  Phys.\ Part.\ Nucl.\  {\bf 29} (1998) 257
   [Fiz.\ Elem.\ Chast.\ Atom.\ Yadra {\bf 29} (1998) 637]
   [Phys.\ Atom.\ Nucl.\  {\bf 61} (1998) 1107]
  [hep-ph/9701269].
\bibitem{B19}
  K.~Agashe, D.~Kim, M.~Toharia and D.~G.~E.~Walker,
  Phys.\ Rev.\ D {\bf 82} (2010) 015007
  [arXiv:1003.0899 [hep-ph]].
\bibitem{B20}
  D.~G.~E.~Walker,
  arXiv:1202.2348 [hep-ph].
\bibitem{FRS1}
R.~Friedberg, T.~D.~Lee and A.~Sirlin,
  Phys.\ Rev.\  D {\bf 13}, 2739 (1976).

\bibitem{FRS2}
R.~Friedberg, T.~D.~Lee and A.~Sirlin,
  Nucl.\ Phys.\  B {\bf 115}, 1 (1976).

\bibitem{FRS25}
R.~Friedberg, T.~D.~Lee and A.~Sirlin,
Nucl.\ Phys.\  B {\bf 115}, 32 (1976).

\bibitem{FRS3}
T.~D.~Lee and Y.~Pang,
  Phys.\ Rept.\  {\bf 221}, 251 (1992).


\bibitem{coleman}
  S.~R.~Coleman,
  Nucl.\ Phys.\ B {\bf 262} (1985) 263
   [Erratum-ibid.\ B {\bf 269} (1986) 744].

\bibitem{susy1}
  A.~Kusenko and M.~E.~Shaposhnikov,
  Phys.\ Lett.\ B {\bf 418} (1998) 46
  [hep-ph/9709492].
\bibitem{susy2}
  A.~Kusenko,
  hep-ph/0001173.



\bibitem{HotBB}
Gorbunov, D.S. and Rubakov, V.A.,
Introduction to the Theory of the Early Universe: Hot Big Bang Theory,
World Scientific, 2011
\bibitem{Guth-Weinberg}
  A.~H.~Guth and E.~J.~Weinberg,
  Phys.\ Rev.\ D {\bf 23} (1981) 876.


\bibitem{Rubakov:1996vz}
  V.~A.~Rubakov and M.~E.~Shaposhnikov,
  Usp.\ Fiz.\ Nauk {\bf 166} (1996) 493
   [Phys.\ Usp.\  {\bf 39} (1996) 461]
  [hep-ph/9603208].

\bibitem{shap1}
A.~Kusenko,~L. C. Loveridge and M. Shaposhnikov, JCAP 0508, 011 (2005) [arXiv:astro-ph/0507225]. 

\bibitem{shap2}
A. Kusenko, L. Loveridge and M. Shaposhnikov, Phys. Rev. D 72, 025015 (2005) [arXiv:hep-ph/0405044]

\bibitem{T}
A. Kusenko, M. E. Shaposhnikov, P. G. Tinyakov and I. I. Tkachev, Phys. Lett. B 423, 104 (1998) [arXiv:hep-ph/9801212]

\bibitem{phtr}
J. R. Espinosa, T. Konstandin, F. Riva, arXiv:1107.5441 [hep-ph]

\bibitem{demgor}
S. V. Demidov, D. S. Gorbunov, JHEP 0702:055,2007, [arXiv:hep-ph/0612368v2]
\end{thebibliography}
\end{document}